\documentclass[amssymb,prl,twocolumn,showpacs]{revtex4}
\input{psfig.sty}
\usepackage{epsfig}
\usepackage{dcolumn}
\usepackage{amsmath}
\usepackage{epsfig}
\begin{document}

\title{Nonequilibrium spintronic transport through an artificial Kondo impurity:
Conductance, magnetoresistance and shot noise} 
\author{Rosa L\'opez}
\author{David S\'anchez}
\affiliation{D\'epartement de Physique Th\'eorique,
Universit\'e de Gen\`eve, CH-1211 Gen\`eve 4, Switzerland}
\date{\today}

\begin{abstract}
We investigate the nonequilibrum transport properties of a quantum dot
when spin flip processes compete with the formation of a Kondo resonance
in the presence of ferromagnetic leads.
Based upon the Anderson Hamiltonian in the strongly interacting limit,
we predict a splitting of the differential conductance when the spin flip
scattering amplitude is of the order of the Kondo temperature.
We discuss how the relative orientation of the lead magnetizations
strongly influences the electronic current and the shot noise
in a nontrivial way. Furthermore, we find that the zero-bias
tunneling magnetoresistance becomes negative with increasing spin flip
scattering amplitude.
\end{abstract}

\pacs{72.15.Qm, 72.25.Mk, 73.63.Kv}
\maketitle

\emph{Introduction}.---The Kondo effect in quantum dots (QD's)
occurs because of a strong antiferromagnetic coupling
between the conduction band electrons in the electrodes and the localized
spin in the QD through higher-order tunneling processes.
The resulting correlated motion gives rise to a Kondo singularity
in the quasiparticle density of states at the Fermi level $E_F$~\cite{ng88}.
As a consequence, the conductance is enhanced below
the Kondo temperature $T_K$~\cite{gol98}.
The fact that the parameters that define $T_K$
are fully controllable and the ability to apply external fields in QD's
have spurred a good deal of works addressing
nonequilibrium situations~\cite{noneq}.
At the same time, recent advances in the control and manipulation of the
spin degree of freedom in magnetic environments~\cite{wol01}
has paved the way for startling
potentialities in spintronic devices~\cite{fie99}
and quantum computation~\cite{los98}. 
A single spin-1/2 confined in a discrete level reaches the ultimate degree
of miniaturization in semiconductor physics. Thus, it is crucial
to understand the properties of spintronic transport in QD's.
The main focus has so far been to investigate magnetic effects
in QD's in the Coulomb blockade regime~\cite{rud01,sou02}.
In comparison, the issue of spin-dependent
physical effects in the Kondo resonance,
which seems to be much richer, has been relatively unexplored.
In particular, one might ask to what extent the Kondo cloud is
sensitive to the lead magnetization.
Or, is the many-body Kondo state robust against spin flip
processes? Answering these questions
is not only interesting for QD's but also
for similar coherent phenomena
in molecular transistors~\cite{par02}.

\begin{figure}
\centerline{
\epsfig{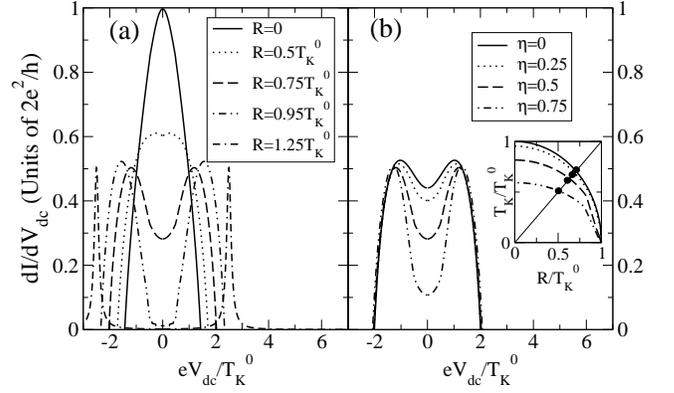}
}
\caption{(a) Differential conductance $\mathcal{G}$ versus bias voltage
$V_{\rm dc}$ in the parallel case with lead magnetizations
$\eta_L=\eta_R=\eta=0.5$
for four different values of the spin flip amplitude $R$.
(b) Dependence of the $\mathcal{G}$--$V_{\rm dc}$ curves on $\eta$
for $R=0.75 T_K^0$.
Inset: Kondo temperature of the QD ($T_K$) as a function of $R$
for the lead polarizations depicted in (b).
The crossing points with the straight line $T_K=R$ marks
the transition to a nonzero splitting in $\mathcal{G}(V_{\rm dc})$.
}
\label{fig1}
\end{figure}

\begin{figure}
\centerline{
\epsfig{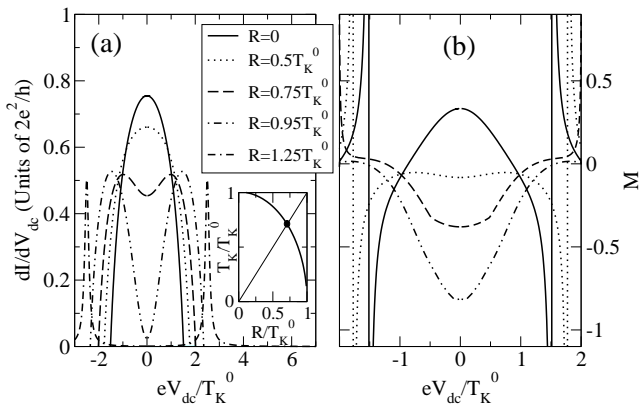}
}
\caption{(a) $\mathcal{G}$--$V_{\rm dc}$ curves in the antiparallel case
for $\eta_L=-\eta_R=0.5$ and different values of $R$.
Inset: $T_K$ as a function of $R$. Notice
that this curve is independent of $\eta$ so that the transition to
the splitting in $\mathcal{G}(V_{\rm dc})$ takes place at
the same point  independently on the value of $\eta$.
(b) Tunneling magnetoresistance $M$ at $\eta=0.5$
as a function of the bias voltage $V_{\rm dc}$
for the first four values of $R$ plotted
in Fig.~\ref{fig2}(a).
}
\label{fig2}
\end{figure}

Consider first a noninteracting QD with a single energy level $\varepsilon_0$
attached to two ferromagnetic contacts. Intradot spin flip processes
lift the level degeneracy, yielding $\varepsilon_0\pm R$,
where $R$ is a phenomenological spin flip scattering amplitude.
When $R$ is greater than the tunneling induced broadening $\Gamma$,
the linear conductance will show a splitting sensitive to
the lead polarization alignment.
Similarly, in the strongly interacting case we expect the
Kondo induced spin fluctuations to become affected by the combined
influence of ferromagnetic leads and spin flip processes. 
Our main findings reflect this physics,
summarized in Figs.~\ref{fig1} and~\ref{fig2}(a):
(i) Let $T_K$ denote
the Kondo temperature of a QD for a given magnetic configuration
($R\neq 0$ and lead magnetization $\eta\neq 0$).
For $R/T_K < 1$ the differential conductance
$\mathcal{G}\equiv dI/dV_{\rm dc}$ ($V_{\rm dc}$ is the bias voltage)
shows the distinctive zero bias anomaly (ZBA)
of Kondo physics~\cite{gol98}
whereas for $R/T_K> 1$
the ZBA smoothly splits at finite
$V_{\rm dc}$ into two peaks separated by a distance $\delta$.
With decreasing $T_K$
(e.g., by lowering $\varepsilon_0$ with a gate voltage),
the spin flip scattering is shown to suppress
the Kondo state. 
(ii) We find that the exact form of $\mathcal{G}$
depends on whether the relative orientation of the lead magnetizations
is parallel (P) or antiparallel (AP).  In addition,
in the P case the ZBA splitting develops at different $R$
with regard to the AP case. We find that in the AP configuration
the variation of the Kondo temperature with $R$ is 
\emph{independent} of $\eta$.
Strong effects of $R$ and $\eta$ in the measurement of
the magnetoresistance and the
current fluctuations at $T=0$ (shot noise) are predicted as well.

\emph{Model}.---There exist in the literature very scarce
theoretical investigations that deal with Kondo transport in QD's in
the presence of ferromagnetic leads~\cite{ser02,mar02,bul03}.
To the best of our knowledge,
no complete picture of the
Fermi-liquid behaviour (i.e., at temperatures $T\ll T_K$) of the Kondo effect
in ultrasmall magnetic tunnel junctions has been yet put forward.
We consider a QD (region 0)
with large on-site Coulomb interaction $U$
coupled via $V_{k\alpha}$ to two Fermi-liquid reservoirs
labeled with $\alpha=\{L,R\}$
with chemical potentials $\mu_L$ and $\mu_R$.
We describe the system with a $N=2$ fold degenerated Anderson
Hamiltonian allowing for intradot spin flips.
In the limit $U\to\infty$, QD double occupancy is forbidden and the 
slave-boson representation~\cite{col84} may be used
to write the Hamiltonian as follows:
\begin{eqnarray}
&&\mathcal{H}=
\sum_{k\alpha\sigma}\varepsilon_{k\alpha\sigma}
c_{k\alpha\sigma}^\dagger c_{k\alpha\sigma}
+\sum_{\sigma}\varepsilon_{0}f_{\sigma}^\dagger f_{\sigma}
+(R f_{\uparrow}^\dagger f_{\downarrow} + {\rm H.c.}) \nonumber\\
&&+\frac{1}{\sqrt{N}}\sum_{k\alpha\sigma} (
V_{k\alpha} c_{k\alpha\sigma}^\dagger b^\dagger f_{\sigma}+
{\rm H.c.} )
+\lambda ( b^\dagger b +\sum_{\sigma} f_{\sigma}^\dagger f_{\sigma} -1) \,,
\label{eq-hamsb}
\end{eqnarray}
where $c_{k\alpha\sigma}^\dagger$ ($c_{k\alpha\sigma}$)
is the creation (annihilation)
operator for an electron in the state $k$ with spin
$\sigma=\{\uparrow,\downarrow\}$ in the lead $\alpha$.
Ferromagnetism in the leads arises through spin-dependent
densities of states $\nu_\sigma$.
We have assumed implicitly that both ferromagnetic
leads possess the same ($z$) easy axis. The generalization to
noncollinear magnetic moments is straightforward~\cite{ser02}.
In Eq.~(\ref{eq-hamsb}), the spin flip term is assumed to be coherent
in the sense that $R$ does not involve spin relaxation
since each flip-flop process can be reversible~\cite{but83}.
Only when $\eta=0$ (no magnetization) do we include in the QD
a vanishingly small Zeeman splitting
to intentionally break the SU(2) spin symmetry
($\varepsilon_{0\uparrow}-\varepsilon_{0\downarrow}=\Delta_Z\to 0^+$).
Finally, $f_{\sigma}^\dagger$ ($f_{\sigma}$) is a pseudofermion operator that
creates (annihilates) a singly occupied state
and the auxiliary boson operator $b^\dagger$ ($b$) creates (annihilates)
an empty state in the QD.
The last term is a constraint enforced by the replacement
of the QD second-quantization operators
by the $f$'s and $b$'s when $U\to\infty$ with an associated Lagrange
multiplier $\lambda$~\cite{hew93}.

The solution of the Hamiltonian~(\ref{eq-hamsb}) can be found in
the mean field approach, which is the leading order in a $1/N$ expansion,
extended to deal with nonequilibrium situations~\cite{agu00}.
This way one sets $b(t)/\sqrt{N}$ to a $c$-number corresponding to its
expectation value $\tilde{b}\equiv \langle b(t)\rangle/\sqrt{N}$,
thereby neglecting charge fluctuations.
This is correct as long as we are interested only in spin fluctuations.
Now, by calculating the equation of motion for $b$
and taking into account the QD constraint we arrive at:
\begin{subequations}\label{eq-sist}
\begin{eqnarray}
\sum_{k\alpha\sigma} \tilde{V}_{k\alpha} G_{\alpha\sigma,0\sigma}^{<} (t,t)
=-iN\lambda |\tilde{b}|^2 \label{eq-b}\,, \\
\sum_{\sigma} G_{0\sigma,0\sigma}^{<} (t,t) = i (1-N|\tilde{b}|^2) \label{eq-q} \,,
\end{eqnarray}
\end{subequations}
which self-consistently determine the unknowns $|\tilde{b}|^2$ and $\lambda$.
We have defined $\tilde{V}_{k\alpha}=|\tilde{b}|^2 V_{k\alpha}$.
The nondiagonal (in the layer indices) Green function
$G_{\alpha\sigma,0\sigma}^{<} (t,t)=
i\langle c_{k\alpha\sigma}^\dagger (t) f_\sigma (t)\rangle$
can be cast in terms of
$G_{0\sigma,0\sigma}^{<} (t,t)=
i\langle f_\sigma^\dagger (t) f_\sigma (t)\rangle$,
with the help of the equation of motion of the operators
and then applying the analytical continuation rules
in a complex time contour~\cite{lan76}.
After lengthy algebra, the Fourier transform of
$G_{0\sigma,0\sigma}^{<}$ becomes
$G_{0\sigma,0\sigma}^{<} (\varepsilon)=2i\sum_\alpha f_\alpha(\varepsilon)
(|M_{\overline{\sigma}}^r|^2 \tilde{\Gamma}_\alpha^\sigma
+R^2\tilde{\Gamma}_\alpha^{\overline{\sigma}})/
|M_\sigma^a M_{\overline{\sigma}}^a-R^2|^2 $,
where $f_\alpha(\varepsilon)$ is the Fermi function
of lead $\alpha$ and
$M_{\overline{\sigma}}^{r,a}=\varepsilon-\tilde{\varepsilon}_{0\sigma}\pm
i\sum_\alpha \tilde{\Gamma}_\alpha^\sigma$.
Notice that by virtue of Kondo correlations,
the original energy level $\varepsilon_{0\sigma}$
is transformed into
$\tilde{\varepsilon}_{0\sigma}=\varepsilon_{0\sigma}+\lambda$.
Likewise, the linewidths $\tilde{\Gamma}_{\alpha}^\sigma=
|\tilde{b}|^2\Gamma_\alpha^\sigma$
renormalize the bare couplings
$\Gamma_\alpha^\sigma (\varepsilon)=(\pi i/2) \sum_{k \alpha} |{V}_{k\alpha}|^2
\delta(\varepsilon-\varepsilon_{k\alpha\sigma})$.
In what follows,
$\Gamma_\alpha^\sigma$ is taken as
$\Gamma_\alpha^\sigma=\Gamma_\alpha^\sigma(E_F)$
for $-D\leq\varepsilon\leq D$
($D$ is the energy cutoff).
Within the slave-boson mean-field theory, Kondo physics arises
when the auxiliary boson field $b$ is condensed.
At that point,
$\tilde{\Gamma}_{\alpha}^\sigma$ gives roughly $T_K$ and
$\lambda$ shifts the resonant level up to $E_F$.

The magnetization at reservoir $\alpha$ is
$\eta_\alpha=(\nu_\alpha^\uparrow-\nu_\alpha^\downarrow)/
(\nu_\alpha^\uparrow+\nu_\alpha^\downarrow)$.
As $-1\le\eta\le 1$, the linewidths $\tilde{\Gamma}_{\alpha}^\sigma$
will be generally spin dependent:
$\Gamma_\alpha^{\uparrow,\downarrow}=(1\pm\eta_\alpha)\Gamma_\alpha/2$.
Hereafter we shall deal with symmetric leads so that
$\Gamma_L=\Gamma_R=\Gamma/2$.
Notice that spin flip effects are fully included
due to the presence of $R$ in the denominator of
$G_{0\sigma,0\sigma}^{<} (\varepsilon)$. 
Hence, the closed form of $G_{0\sigma,0\sigma}^{<} (\varepsilon)$
given above represents
the distribution function of a QD in the
presence of spin flips, ferromagnetic leads and Zeeman splitting~\cite{note}.
For $R=0$ and $\eta_\alpha=0$
we get at equilibrium the position of the Kondo resonance of the impurity at
$\tilde{\varepsilon}_{0}=\tilde{\varepsilon}_{0\uparrow}=
\tilde\varepsilon_{0\downarrow}$
with width $\tilde{\Gamma}_\alpha= \tilde{\Gamma}_\alpha^\downarrow+
\tilde{\Gamma}_\alpha^\uparrow$ and Kondo temperature
$T_K^0=(\tilde{\varepsilon}_{0}^2+\tilde{\Gamma}^2)^{1/2}=
D\exp{(-\pi|\varepsilon_0|/\Gamma)}$.
From the solution of Eq.~(\ref{eq-sist}), we can obtain the
current traversing the dot:
$I=(-e/\pi\hbar) {\rm Im} \sum_\sigma \tilde{\Gamma}_L^\sigma
\int [2G_{0\sigma,0\sigma}^{r}(\varepsilon)
f_L(\varepsilon)+G_{0\sigma,0\sigma}^{<}(\varepsilon)] d\varepsilon$,
$f(E)$ being the Fermi function. In the following, we shall present
mainly \emph{nonequilibrium} results for 
$\mu_L=-\mu_R=eV_{\rm dc}/2$, $\varepsilon_0=-3.5\Gamma$ and $D=-60\Gamma$
($T_K^0\approx 10^{-3}\Gamma$). The reference energy is set at $E_F=0$.

\emph{Results}.---
Let us first focus on the case of polarized reservoirs with parallel (P)
alignment ($\eta_L=\eta_R=\eta$).
Figure~\ref{fig1}(a) shows $\mathcal{G}$ for partial lead magnetization ($\eta=0.5$) and different values of $R$. 
As seen, $\mathcal{G}$ is a \emph{direct} measure of how the spin flip 
mechanisms weaken the Kondo effect. For relatively small values of $R$ ($R<T_K$) the height of the ZBA decreases but $\mathcal{G}$ does not split. The appearance of a splitting $\delta\neq 0$ occurs at $R\sim T_K$.  Importantly, due to that coexistence of the Kondo state and
the intradot spin flips, the positions of the peaks in $\mathcal{G}$
are renormalized by Kondo correlations, i.e.,
these peak positions are \emph{not} trivially related with
their single-particle counterparts
(which should be centered around $V_{dc}=\pm 2R$). For $R>T_K^0$, the mean field approach predicts a complete quenching of the ZBA ($T_K=0$) because the solution of Eq.~(\ref{eq-sist}) is the trivial one ($\tilde{b}=0$), which is unphysical. A more detailed theory including charge fluctuations
(noncrossing approximation) would correct this.
They price that one has to pay, however,
is the appearance of spurious peaks in the conductance.  Only for illustrative purposes have we also plotted in Fig.~\ref{fig1}(a) the case with $R=1.25 T_K^0$~\cite{note_r}. Here, $\mathcal{G}$ is maximal at $V_{dc}\sim\pm 2R$.
Notice that for $R>T_K^0$ our model overestimates the
height of the peaks as they should quickly tend to
zero in the Coulomb blockade valley.

In Fig.~\ref{fig1}(b) we plot the dependence
of $\mathcal{G}$ on the lead polarization (for P alignment)
at a fixed value of $R$.
The overall effect of $\eta$ is to diminish $\mathcal{G}$.
In particular, the ZBA quenching can be ascribed to a weaker Kondo
state because of a smaller $T_K$.
The Kondo temperature can be easily obtained from Eq.~(\ref{eq-sist})~\cite{col84}
but for $R\neq 0$ and $\eta\neq 0$ we find an implicit equation that
can be worked out numerically (in contrast to the AP case, see below).
We depict the results in the
inset of Fig.~\ref{fig1}(b).
The splitting $\delta\neq 0$ takes place
at a lower $R$ upon increasing $\eta$, in agreement
with Fig.~\ref{fig1}(b).
We have checked that our numerical results
fulfill two requirements, namely: (i) 
they are independent of the sign of $\eta$, for $T_K$ must
be invariant after flipping simultaneously
all the electron spins in both reservoirs, and (ii)
with increasing $\eta$,
$T_K$ decreases [see inset in Fig.~\ref{fig1}(b) along the vertical axis $R=0$]
and eventually vanishes for full lead polarization ($\eta\to 1$),
i.e., no Kondo effect
may arise in the case of half-metallic leads.

\begin{figure}
\centerline{
\epsfig{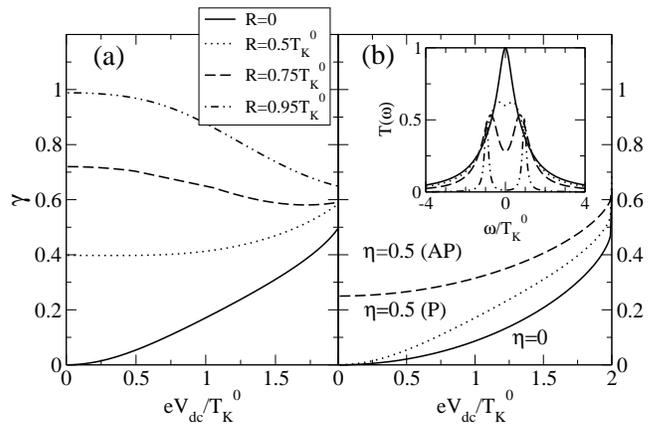}
}
\caption{Fano factor versus bias voltage as a function of $R$
for the parallel alignment with $\eta_L=\eta_R=0.5$.
(b) Voltage dependence of the shot noise for the unpolarized,
parallel and antiparallel case at $R=0$. Inset: transmission
probability as a function of energy (the origin is taken at $E_F=0$)
for the same values of $R$ as in (a).
}
\label{fig3}
\end{figure}

Unlike the P case, for the AP alignment ($\eta_L=-\eta_R=\eta$)
the ZBA splitting is generated at larger values of $R$
[see Fig.~\ref{fig3}(a) for $\eta=0.5$].
Nevertheless, we observe that the ZBA height is reduced even for $R=0$.
The reason is that $\mathcal{G}$ decreases roughly by a factor $1-\eta^2$
when the magnetic moments point along the opposite directions.
More interesting is the fact that the ZBA width is given by
a $|\eta|$-independent Kondo temperature.
We find $T_K=\sqrt{(T_K^0)^2-R^2}$, i.e., $T_K$
does not depend on $|\eta|$ [see inset of Fig.~\ref{fig2}(a)].
The underlying physics for this is that
both spin channels are coupled to the QD in the same way
(i.e., $\Gamma_L^\uparrow+\Gamma_R^\uparrow=
\Gamma_L^\downarrow+\Gamma_L^\downarrow=\Gamma$) so that $T_K$ persists unaltered \emph{regardless of the value of $\eta$}. As a result, the splitting in $\mathcal{G}(V_{\rm dc})$ takes place always at the same point independently on the degree of lead magnetization.  

We have found as well interesting features in the
tunneling magnetoresistance
$M\equiv (\mathcal{G}^{\rm P}-\mathcal{G}^{\rm AP})/\mathcal{G}^{\rm AP}$
[see Fig.~\ref{fig3}(b)]. For definiteness,
let us focus on the $V_{\rm dc}=0$ case.
In conventional magnetic tunnel junctions, it is usual that the
P current is larger than the AP current, thereby resulting in an increase
of the impedance of the system when the configuration
is switched from the P to the AP alignment~\cite{pri98}.
This is really the case for the Kondo ZBA in the absence of
spin flips [see Fig.~\ref{fig2}(b) for $R=0$]. However, increasing $R$
does not lead only to a reduction of $M$
but also to a \emph{reversal} of the sign of $M$.
This arises from the fact that the Kondo effect is more robust against
spin flips in the AP configuration than in the P alignment, as stated above.

\emph{Shot noise}.--- As known,
current fluctuations due to charge granularity (shot noise)
can provide information additional to the averaged current~\cite{bla02}.
The shot noise is given by the current-current
correlation function $S(t-t')=\langle \delta \hat{I}(t)
\delta \hat{I}(t') \rangle$,
where $\delta \hat{I} =\hat{I} - \langle \hat{I} \rangle$
measures the current fluctuations.
(Because we consider a two-terminal system, we have dropped
the lead indices.)
An important figure of merit is the Fano factor:
$\gamma=P(0)/2e\langle I\rangle$, where
$P(0)$ is the noise power spectrum at zero-frequency
[i.e., the Fourier transform of $S(t-t')$].
The Fano factor determines the deviations of the shot noise away
from the Poissonian value of a classical conductor due to, e.g.,
strong electron-electron interactions
as those giving rise to the Kondo effect.
Figure~\ref{fig3} shows the influence of
spin flips (a) and lead magnetization (b) in $\gamma$.
At low bias, $\gamma$ behaves
as $1-\mathcal{T}(E_F)$~\cite{bla02},
where $\mathcal{T}$ is the transmission probability. 
As shown in Fig.~\ref{fig3}(a) for $R=0$, the Kondo unitary limit is reached for a partial lead magnetization in the parallel case.  Because of the correlated motion of the electrons which lead to the singlet formation there is a suppression of $\gamma$ at zero bias. As $R$ is turned on, the transmission departs from its unitary limit [see inset of Fig.~\ref{fig3}(b)] because the Kondo effect is quenched and $\gamma$ takes on a nonzero value. Finally, for the largest $R$ the Fano factor tends to its Poissonian limit as spin flip processes reduce the Kondo resonance and a two-peak structure arise in the transmission probability. Thus, while the splitting of $\mathcal{G}$ requires large applied voltages to be observable, the shot noise shows the same effect already in the \emph{zero} voltage limit.
On the other hand,
the behavior of $\gamma$ at larger $V_{\rm dc}$ depends on the specific
transport mechanism. At $R=0$, correlations between the band electrons and
the localized electron
dominate so that $\gamma$ increases with $V_{\rm dc}$~\cite{don02}.
For $R/T_K^0\to 1$, the Kondo resonance is split [see inset in Fig.~\ref{fig3}(b)] 
 and transport is maximized
as $V_{\rm dc}$ sweeps across the new resonances.
For comparison between normal metals and ferromagnetic electrodes, we observe from Fig.~\ref{fig3}(b) ($R=0$) that (i) $\gamma$ increases more rapidly for the P than for the unpolarized case due to the $\eta$ dependence of $T_K$,
and (ii) for AP alignment the ZBA diminishes by a factor $1-\eta^2$
and therefore we find $\gamma\sim \eta^2$ at low bias.

\emph{Conclusion}.---Using a slave-boson mean-field theory,
we have shown that the combined influence of
ferromagnetic electrodes and spin flip transitions
in the Kondo physics of a QD
manifests itself in the nonequilibrium transport properties of the system.
The Kondo temperature (which can be regarded as the coupling strength
of the singlet formation) is shown to be suppressed with increasing
lead magnetization in the parallel alignment but the Kondo state is
remarkably stable in the antiparallel case. This may lead to
negative tunneling magnetoresistance at zero-bias when the spin flip
scattering amplitude is enhanced. 
The overall behavior is confirmed with shot noise calculations.
Experimentally, we believe that the effects addressed in this paper
should be visible within the scope
of present techniques as our energies are within the Kondo scale.

We gratefully acknowledge R. Aguado, M. B\"uttiker, G. Platero and
E.V. Sukhorukov for helpful discussions.
Work supported by the EU
RTN under contract HPRN-CT-2000-00144, Nanoscale Dynamics.


\begin{thebibliography}{90}
\bibitem{ng88}
T.K. Ng and P.A. Lee, Phys. Rev. Lett. {\bf 61}, 1768 (1988);
L.I. Glazman and M.E. Raikh, JETP Lett. {\bf 47}, 452 (1988).
\bibitem{gol98}
D. Goldhaber-Gordon {\it et al.}, Nature (London) {\bf 391}, 156 (1998);
S.M. Cronenwett {\it et al.}, Science {\bf 281}, 540 (1998);
J. Schmid {\it et al.}, Physica B {\bf 256-258}, 182 (1998).
\bibitem{noneq}
S. Hershfield {\it et al.}, Phys. Rev. Lett. {\bf 67}, 3720 (1991);
Y. Meir {\it et al.}, {\it ibid.} {\bf 70}, 2601 (1993);
N.S. Wingreen and Y. Meir, Phys. Rev. B {\bf 49}, 11040 (1994);
R. L\'opez {\it et al.}, Phys. Rev. Lett. {\bf 81}, 4688 (1998);
A. Kaminski {\it et al.}, {\it ibid.} {\bf 83}, 384 (1999);
M. Plihal {\it et al.}, Phys. Rev. B {\bf 61}, R13341 (2000);
P. Coleman {\it et al.},  Phys. Rev. Lett. {\bf 86}, 4088 (2001).
\bibitem{wol01}
See, e.g., S.A. Wolf {\it et al.},
Science {\bf 294}, 1488 (2001) and references therein.
\bibitem{fie99}
R. Fiederling {\it et al.}, Nature (London) {\bf 402}, 787 (1999);
Y. Ohno {\it et al.}, {\it ibid.} {\bf 402}, 790 (1999).
\bibitem{los98}
D. Loss and D.P. DiVincenzo, Phys. Rev. A {\bf 57}, 120 (1998);
P. Recher {\it et al.}, Phys. Rev. Lett. {\bf 85}, 1962 (2000).
\bibitem{rud01}
W. Rudzi\'nski and J. Barna\'s, Phys. Rev. B {\bf 64}, 085318 (2001).
\bibitem{sou02}
F.M. Souza {\it et al.}, cond-mat/0209263 (unpublished).
\bibitem{par02}
J. Park {\it et al.}, Nature (London) {\bf 417}, 722 (2002);
W. Liang {\it et al.}, {\it ibid.} {\bf 417}, 725 (2002).
\bibitem{ser02}
N. Sergueev {\it et al.}, Phys. Rev. B {\bf 65}, 165303 (2002).
\bibitem{mar02}
J. Martinek {\it et al.}, cond-mat/0210006 (unpublished).
\bibitem{bul03}
B.R. Bulka and S. Lipi\'nski, Phys. Rev. B {\bf 67}, 024404 (2003).
\bibitem{col84}
P. Coleman, Phys. Rev. B {\bf 29}, 3035 (1984).
\bibitem{but83}
In fact, $R$ induces a Larmor precession around an axis perpendicular
to the $z$-axis; see M. B\"uttiker, Phys. Rev. B {\bf 27}, 6178 (1983).
\bibitem{hew93}
A.C. Hewson, {\it The Kondo Problem to Heavy Fermions}
(Cambridge University Press, Cambridge, UK, 1993).
\bibitem{agu00}
R. Aguado and D.C. Langreth, Phys. Rev. Lett. {\bf 85}, 1946 (2000).
\bibitem{lan76}
D.C. Langreth, in \emph{Linear and Nonlinear Electron Transport in Solids}
(J.T. Devreese and V.E. Van Doren, eds.),
NATO ASI, Ser. B, Vol. 17 (Plenum, New York, 1976). 
\bibitem{note} In the slave-boson mean-field theory,
the spin and charge fluctuations
are fully decoupled. This means in particular that we cannot detect
possible lead magnetization induced Kondo peak splittings, as those
predicted in Ref.~\cite{mar02} for the parallel configuration,
unless $1/N$ corrections are included in the theory.
Although work in this direction is being made, here we are just interested
in a good description of spin induced effects
around a well defined Kondo resonance, thereby ignoring charge
fluctuations. In fact, we have checked that the splitting vanishes
in this regime (for very low $T$ and deep $\varepsilon_{0}$).
\bibitem{note_r} Here, the peak width is put by hand as
the slave-boson mean-field theory
would yield zero width. Yet, in an actual QD this
divergence would be regularized by charge fluctuations.
\bibitem{pri98}
G.A. Prinz, Science {\bf 282}, 1660 (1998).
\bibitem{bla02}
For a review, see Ya. M. Blanter and M. B\"uttiker, Phys. Rep. {\bf 336}, 2 (2002).
\bibitem{don02}
Y. Meir and A. Golub, Phys. Rev. Lett. {\bf 88}, 116802 (2002); 
B. Dong and X.L. Lei, J. Phys.: Condens. Matter {\bf 14}, 4963 (2002).
\end{thebibliography}
\end{document}